# High Curie temperature $Ga_{1-x}Mn_xAs$ obtained by resistance-monitored annealing

K.W. Edmonds, K.Y. Wang, R.P. Campion, A.C. Neumann, N.R.S. Farley, B.L. Gallagher, and C.T. Foxon

*School of Physics and Astronomy, University of Nottingham, Nottingham NG7 2RD, UK*

*We show that by annealing $Ga_{1-x}Mn_xAs$ thin films at temperatures significantly lower than in previous studies, and monitoring the resistivity during growth, an unprecedented high Curie temperature $T_C$ and conductivity can be obtained. $T_C$ is unambiguously determined to be 118 K for Mn concentration x=0.05, 140 K for x=0.06, and 120 K for x=0.08. We also identify a clear correlation between $T_C$ and the room temperature conductivity. The results indicate that Curie temperatures significantly in excess of the current values are achievable with improvements in growth and post-growth annealing conditions.*

The p-type dilute magnetic semiconductor $Ga_{1-x}Mn_xAs$ is potentially of great importance for magnetoelectronic applications. However, the realisation of this will require a substantial increase in the ferromagnetic transition temperature $T_C$. The current record value of ≈110 K [1] was obtained soon after the first discovery of ferromagnetism in this system [2]. More recently, post-growth annealing was identified as a technique for increasing $T_C$. Hayashi *et al.* [3] found that for short annealing times (15 minutes), the largest $T_C$ values were obtained for annealing temperatures slightly in excess of the growth temperature, while elsewhere [4,5] a 90 minute anneal at the growth temperature was reported to optimise $T_C$. Longer annealing times and higher temperatures resulted in a reduction of $T_C$ [3-6]. Yu *et al.* [7] showed strong evidence that the increase of $T_C$ was related to the removal of Mn interstitial atoms that occur during the non-equilibrium growth of $Ga_{1-x}Mn_xAs$, and that act as double donors, compensating the itinerant holes and thus reducing the carrier mediated ferromagnetism. In all cases, the improved $T_C$ is accompanied by an increased hole density and conductivity, consistent with the removal of compensating defects. However, in spite of a claim that the annealing procedure employed in ref. [5] is optimised, the $T_C$ values obtained do not exceed 110 K. This is in contrast to mean field predictions of room temperature ferromagnetism in $Ga_{0.9}Mn_{0.1}As$ [8], leading to suggestions that $T_C$ in $Ga_{1-x}Mn_xAs$ is fundamentally limited to 110 K [7].

Here we show that, by carefully controlling the growth and annealing conditions, Curie temperatures significantly in excess of 110 K can be obtained. In contrast to the earlier studies, much longer annealing times and lower temperatures are employed. Furthermore, the effect of the anneal is monitored *in-situ* by measurement of the resistivity. The resistivity is shown to give a good measure of the effect of annealing on $T_C$.

45nm thick $Ga_{1-x}Mn_xAs$ samples with 0.015≤ *x* ≤0.08 were prepared on GaAs(001) substrates by low-temperature molecular beam epitaxy using $As_2$. The samples were grown at the highest temperatures possible while maintaining two-dimensional growth; the growth temperature $T_g$ for the samples discussed is shown in table I. However, we note that the quantitative value of $T_g$ cannot be measured with a high degree of accuracy, and therefore cannot be directly compared with the post-growth annealing temperature. The use of $As_2$ in preference to $As_4$ has been shown to lower the concentration of As antisite defects, so that the Mn interstitial is likely to be the most important compensating defect [9]. Therefore, we expect low-temperature annealing to have a pronounced effect on these films. The as-grown samples showed metallic conduction at low temperatures across the whole doping range. The growth, structure, and magneto-transport of these samples is discussed in detail elsewhere [9,10]. The samples were photo-lithographically fabricated into Hall bars, and the surface was capped with PMMA to prevent oxidation and As desorption during the anneal. Annealing was performed in an oven at a temperature of (175±5)ºC, while monitoring the resistance of the Hall bars using an ac resistance bridge. The effect of annealing on $T_C$ is expected to be most pronounced at high Mn concentrations, as reported elsewhere [5]. Therefore, we focus our study on samples with *x*=0.05-0.08.

Figure 1 shows the resistivity $\rho$ of samples with *x*=0.05 and *x*=0.06, monitored *in-situ* while annealing, as a function of the annealing time. We have established that the sample comes into equilibrium at the anneal temperature in less then one hour. For both samples, $\rho$ falls monotonically throughout the anneal, with a rate that decreases with increasing time. This is in contrast to ref. [4], where for significantly higher annealing temperatures the resistivity (measured at room temperature after annealing) was found to fall initially, before increasing for annealing times exceeding 2 hours, with a similar trend observed for $T_C$.

The increased resistance observed previously for long annealing times suggests the presence of two competing thermally activated processes, giving respectively an increase and a decrease in the conductivity and $T_C$. The first mechanism was originally ascribed to the diffusion of As antisite defects [4]; however, recent evidence has shown that the annealing-enhanced $T_C$ is related to the removal of highly mobile Mn interstitial atoms, for which the activation energy is relatively low [7]. The reduction of $T_C$ at longer annealing times may be due to the removal of Mn from the electrically active Ga sites, for which the activation energy will be much higher. Therefore, the much lower annealing temperature employed in the present study compared with elsewhere (175ºC compared with 250ºC [4] or 282ºC [7]) leads to an almost complete suppression of the latter mechanism while still allowing for the precipitation of Mn interstitials (albeit on a longer timescale).

The low temperature annealing has a pronounced effect on $T_C$ and the conductivity, as is shown in figure 2. Figure 2a shows $\rho$ versus temperature $T$ for the sample with $x=0.06$, as-grown and after a 230 hour anneal at 180ºC. Both curves show typical behaviour for metallic $Ga_{1-x}Mn_xAs$ [1], with $\rho$ showing an initial increase on cooling from room temperature, followed by a peak corresponding approximately to $T_C$ and a sharp decrease, and finally increasing again for temperatures below around 4 K. The reduction of $\rho$ due to annealing becomes larger with decreasing temperature. Also, the peak in the $\rho$ versus $T$ curve moves from around 80 K in the as-grown sample to 165 K in the annealed sample, signifying an increase in $T_C$.

The peak in the $\rho$ versus $T$ curve gives a reasonable estimate of $T_C$ for relatively low values ($T_C$<100 K), within the errors associated with the broadness of the peak, but is found to overestimate $T_C$ for higher values. A more accurate method of obtaining $T_C$ is by using the anomalous Hall effect. This is large in $Ga_{1-x}Mn_xAs$, dominating the normal Hall effect even at temperatures much higher than $T_C$ [1,6], and is proportional to the perpendicular component of the sample magnetisation. The Hall resistance $R_{xy}$ of the annealed 6% sample is shown in figure 2b, for an external magnetic field $B$ applied perpendicular to the sample surface, at sample temperatures 110 K, 130 K, and 140 K.

A complication with determining the temperature dependence of the magnetisation $M(T)$ from these measurements is that the anomalous Hall resistance is also an uncertain function of $\rho$, with skew and side-jump scattering models predicting either a linear or quadratic dependence [1]. However, $T_C$ can be determined with a high degree of accuracy using Arrott plots [1,10]. Here, $M^2$ is plotted as a function of $B/M$ at a given temperature, where $M$ is the magnetisation given by $R_{xy}/\rho^n$, with $n=1$ or 2 depending on the scattering mechanism. The linear part of the graph at high fields is extrapolated to $B/M=0$, and the intercept of this at the y-axis is proportional to the square of the saturation magnetisation. Some Arrott plots for temperatures close to $T_C$ for the annealed $x=0.06$ sample are shown in the inset of figure 2c. These plots are for $n=2$; however, using $n=1$ gives the same value for $T_C$ within an error of ±1 K, so $T_C$ can be determined unambiguously by this method.

$M(T)$ for the annealed $x=0.06$ sample is shown in figure 2c for the $n=1$ and $n=2$ models. We obtain a value of $T_C=140$ K, significantly exceeding the highest values of ≈110 K recorded previously [1,3-6]. Also shown in figure 2c is a Brillouin function for spin 5/2 and $T_C=140$ K. The $n=2$ model gives better agreement with this, in agreement with a recent study of the anomalous Hall effect at high magnetic fields and low temperatures [11].

The as-grown values of $T_C$ and the conductivity, and the highest values obtained by annealing, are shown in table I for the different Mn concentrations. The Zener model of ferromagnetism, which has been successful in describing many of the properties of dilute magnetic semiconductor systems [8,12,13], predicts a $T_C$ that scales approximately as $xp^{1/3}$, where $p$ is the density of holes. Therefore, $T_C$ should be largest when compensation is weak and $p \approx x$, giving $T_C \sim x^{4/3}$. Using this relationship and the parameters given in ref. [8], we obtain theoretical maximum values of $T_C \approx 150$ K, 190 K, and 280 K for $x=0.05$, 0.06, and 0.08 respectively. Thus, although the observed Curie temperatures represent a significant improvement over all previously reported values, they still fall short of these ideal values by a sizeable margin. Although it has been predicted that for low compensation the mean field model breaks down and the carrier-induced ferromagnetism is suppressed due to the increasing importance of antiferromagnetic RKKY oscillations [14], this is not borne out by experiment [10].

The reduction in $T_C$ observed experimentally on going from $x=0.06$ to $x=0.08$ may be related to a higher defect density in the latter sample. The growth temperature is decreased with increasing Mn concentration. Therefore, the lower growth temperature for $x=0.08$ is likely to result in the incorporation of higher numbers of compensating As antisite defects, which are stable at the low annealing temperatures employed in the present study [7]. However, it may be possible to obtain a further increase in $T_C$ with longer anneals or different annealing temperatures.

Figure 3 shows the Curie temperature for a series of $Ga_{1-x}Mn_xAs$ samples, as-grown and annealed, as a function of the conductivity $\sigma_{300K}$ measured at room temperature and under zero magnetic field. For all the samples, $T_C$ increases with increasing $\sigma_{300K}$, with a gradient that increases with increasing Mn concentration. The data points in figure 3 correspond to samples annealed at different temperatures and times, and also in the case of $x=0.08$ to samples taken from different wafers. In spite of this, a clear trend can be observed, indicating that the high-temperature resistivity is a good measure of $T_C$.

Jungwirth *et al.* have calculated zero temperature conductivities in $Ga_{1-x}Mn_xAs$ within the Boltzmann relaxation time approximation, including scattering from the random Mn ion distribution as well as from Mn interstitials and As antisites [13]. For weakly compensated samples with Mn concentration $x$=0.06, theoretical conductivities in excess of 5000 $\Omega^{-1}$cm$^{-1}$ were obtained. The highest conductivity obtained in the present study is considerably less than this prediction. If we assume, as in ref. [13], that the large difference is due to additional defects occurring during growth, which are not removed by the low-temperature annealing process and are not accounted for by the theoretical model, then this implies that there is still considerable scope for increasing the conductivity, with increased control of the growth conditions and post-growth treatment. Therefore, if the trend shown in figure 3 continues to higher conductivities, this will be accompanied by a considerable increase in $T_C$, bringing this closer to the mean-field-predicted values and opening up the possibility of room temperature ferromagnetic $Ga_{1-x}Mn_xAs$.

In summary, by careful control of growth conditions and resistance-monitored low-temperature annealing of $Ga_{1-x}Mn_xAs$, we have obtained ferromagnetic transition temperatures up to 140 K. This conclusively demonstrates that $T_C$ is not fundamentally limited to 110 K. We have no reason to think that our growth and annealing procedures are optimum. These results indicate that further increases are to be expected and that we are not close to some fundamentally limited $T_C$ in this system.

| x | $T_g$ (°C) | as-grown | | | annealed | | |
|---|---|---|---|---|---|---|---|
| | | $T_C$ (K) | $\sigma_{300K}$ ($\Omega^{-1}$cm$^{-1}$) | $\sigma_{4.2K}$ ($\Omega^{-1}$cm$^{-1}$) | $T_C$ (K) | $\sigma_{300K}$ ($\Omega^{-1}$cm$^{-1}$) | $\sigma_{4.2K}$ ($\Omega^{-1}$cm$^{-1}$) |
| 0.05 | 210 | 76 | 294 | 345 | 118 | 478 | 671 |
| 0.06 | 200 | 76 | 270 | 286 | 140 | 531 | 757 |
| 0.08 | 185 | 56 | 135 | 32 | 120 | 313 | 349 |

Table I. Mn concentration $x$, growth temperature $T_g$, Curie temperature $T_C$, conductivity at room temperature $\sigma_{300K}$ and at 4.2K $\sigma_{4.2K}$, for as-grown and annealed Ga$_{1-x}$Mn$_x$As thin films.

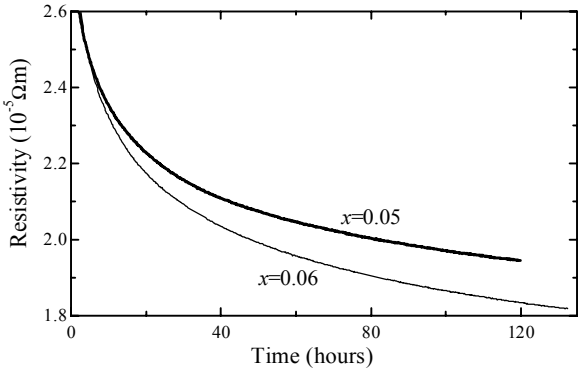

Figure 1: *in-situ* monitored resistivity versus annealing time at 175°C, for GaMnAs thin films with 5% (thick line) and 6% (thin line) Mn concentrations.

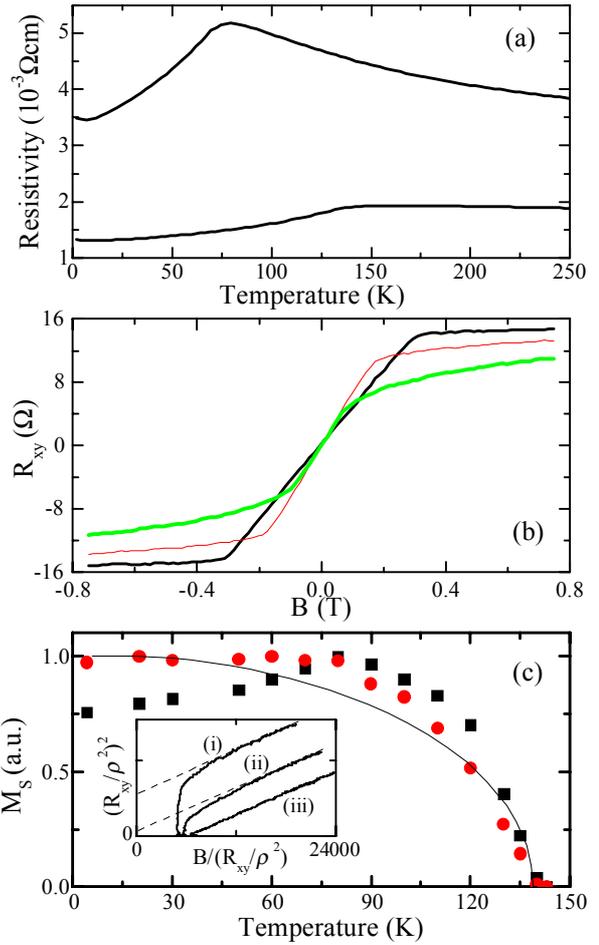

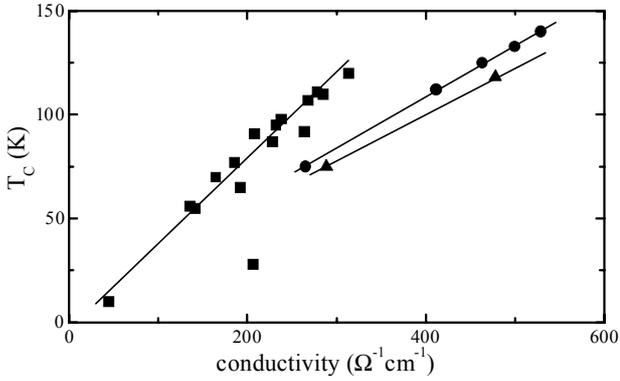

Figure 3: Curie temperature versus room temperature conductivity for GaMnAs films with $x$=0.08 (squares), $x$=0.06 (circles), $x$=0.05 (triangles). Straight lines are to guide the eye.

Figure 2: (a) resistivity versus temperature for as-grown and annealed Ga$_{0.94}$Mn$_{0.06}$As; (b) Hall resistance versus external magnetic field for annealed Ga$_{0.94}$Mn$_{0.06}$As at 110 K (black), 130 K (red), 140K (green); (c) saturation magnetisation versus temperature obtained from Arrott plots for annealed Ga$_{0.94}$Mn$_{0.06}$As assuming linear (open squares) and quadratic (closed squares) dependence of anomalous Hall resistance on $\rho$, and Brillouin function for $T_C$=140 K (line); inset of (c) Arrott plots at (i) 135 K, (ii) 140 K, (iii) 143 K.